\documentclass[12pt,english]{article}
\usepackage[T1]{fontenc}
\usepackage[latin9]{inputenc}
\usepackage{geometry}
\usepackage{amsmath}
\usepackage{setspace}
\usepackage{amssymb}

\makeatletter
\newcommand{\lyxaddress}[1]{
\par {\raggedright #1
\vspace{1.4em}
\noindent\par}
}

\usepackage{babel}
\makeatother

\begin{document}

\title{Quaternion Octonion Reformulation of Quantum Chromodynamics}

\author{Pushpa$^{(1)}$, P. S. Bisht$^{(1)}$, Tianjun Li$^{\text{(2)}}$,
and O. P. S. Negi$^{(1)}$}

\maketitle
\begin{center}

\par\end{center}

\lyxaddress{\begin{center}
$^{(1)}$Department of Physics,\\
 Kumaun University, S. S. J. Campus, \\
Almora-263601 (Uttarakhand) India
\par\end{center}}

\lyxaddress{\begin{center}
$^{(2)}$Institute of Theoretical Physics,\\
 Chinese Academy of Sciences,\\
 P. O. Box 2735,\\
 Beijing 100080, P. R. China 
\par\end{center}}

\lyxaddress{\begin{center}
Email-pushpakalauni60@yahoo.co.in,\\
 ps\_bisht 123@rediffmail.com, \\
tli@itp.ac.cn, \\
ops\_negi@yahoo.co.in.
\par\end{center}}

\begin{abstract}
We have made an attempt to develop the quaternionic formulation of
Yang \textendash{} Mill\textquoteright{}s field equations and octonion
reformulation of quantum chromo dynamics (QCD). Starting with the
Lagrangian density, we have discussed the field equations of $SU(2)$
and $SU(3)$ gauge fields for both cases of global and local gauge
symmetries. It has been shown that the three quaternion units explain
the structure of Yang- Mill\textquoteright{}s field while the seven
octonion units provide the consistent structure of $SU(3)_{C}$ gauge
symmetry of quantum chromo dynamics.

\textbf{Key Words:} Quaternion, Octonions, Quantum Chromodynamics

\textbf{PACS No}.: 14.80 Hv.
\end{abstract}

\section{Introduction}

~~~~~~~The role of number system ( hyper complex number ) is
an important factor for understanding the various theories of physics
from macroscopic to microscopic level. In elementary particle physics
, electromagnetism , the strong and weak nuclear forces are described
by a combination of relativity and quantum mechanics called relativistic
quantum field theory. The electroweak and strong interactions are
described by the Standard Model (SM). Standard Model unifies the Glashow
\textendash{} Salam \textendash{} Weinberg (GSW) electroweak theory
and the quantum chromodynamics (QCD) theory of strong interactions.
According to celebrated Hurwitz theorem \cite{key-1} there exits
four-division algebra consisting of $\mathbb{R}$ (real numbers),
$\mathbb{C}$ (complex numbers), $\mathbb{H}$ (quaternions) \cite{key-2,key-3}
and $\mathcal{O}$ (octonions) \cite{key-4,key-5,key-6}. All four
algebras are alternative with antisymmetric associators. Real number
explains will the classical Newtonian mechanics,complex number play
an important role for the explanation beyond the framework of quantum
theory and relativity. Quaternions are having relations with Pauli
matrices explain non abelian gauge theory. Quaternions were very first
example of hyper complex numbers having the significant impacts on
mathematics \& physics. Because of their beautiful and unique properties
quaternions attracted many to study the laws of nature over the field
of these numbers. Yet another complex system i.e; Octonion may play
an important role \cite{key-6,key-7,key-8,key-9} in understanding
the physics beyond strong interaction between color degree of freedom
of quarks and their interaction. Quaternions naturally unify \cite{key-10}
electromagnetism and weak force, producing the electroweak $SU(2)\times U(1)$
sector of standard model. Octonions are used for unification programme
for strong interaction with successful gauge theory of fundamental
interaction i.e; octonions naturally unify \cite{key-11} electromagnetism
and weak force producing $SU(3)_{c}\times SU(2)_{w}\times U\,(1)_{Y}.$
In this paper, we have made an attempt to develop the quaternionic
formulation of Yang \textendash{} Mill\textquoteright{}s field equations
and octonion reformulation of quantum chromo dynamics (QCD). Starting
with the Lagrangian density, we have discussed the field equations
of $SU(2)$ and $SU(3)$ gauge symmetries in terms of quaternions
and octonions. It has been shown that the three quaternion units explain
the structure of Yang- Mill\textquoteright{}s field while the seven
octonion units provide the consistent structure of $SU(3)_{C}$ gauge
symmetry of quantum chromodynamics (QCD) as they have connected with
the well known $SU(3)$ Gellmann $\lambda$ matrices. In this case
the gauge fields describe the potential and currents associated with
the generalized fields of dyons particles carrying simultaneously
the electric and magnetic charges.

\section{Quaternionic Lagrangian Formulation}

Let us consider that we have two spin $1/2$ fields, $\psi_{a}$ and
$\psi_{b}$. The Lagrangian without any interaction is thus defined
\cite{key-12} as \begin{eqnarray}
L=[i\overline{\psi}_{a}\gamma^{\mu}\partial_{\mu}\psi_{a}-m\overline{\psi}_{a}\psi_{a}]+[ & i\overline{\psi}_{b}\gamma^{\mu}\partial_{\mu}\psi_{b}-m\overline{\psi}_{b}\psi_{b} & ]\label{eq:1}\end{eqnarray}
where $m$ is the mass of particle, $\overline{\psi}_{a}$ and $\overline{\psi}_{b}$
are respectively used for the adjoint representations of $\psi_{a}$
and $\psi_{b}$ and the $\gamma$ matrices are defind as 

\begin{equation}
\gamma_{0}=\left(\begin{array}{cc}
1 & 0\\
0 & -1\end{array}\right);\,\,\,\,\,\,\,\,\,\gamma_{j}=\left(\begin{array}{cc}
0 & \sigma_{j}\\
-\sigma_{j} & 0\end{array}\right)(\forall\, j=1,2,3).\label{eq:2}\end{equation}
Here $\sigma_{j}$ are the well known $2\times2$ Pauli spin matrices.
Lagrangian density (\ref{eq:1}) is thus the sum of two Lagrangians
for particles $a$ and $b$. We can write above equation more compactly
by combining $\psi_{a}$ and $\psi_{b}$ into two component column
vector;\begin{eqnarray}
\psi & = & \left(\begin{array}{c}
\psi_{a}\\
\psi_{b}\end{array}\right)\label{eq:3}\end{eqnarray}
and accordingly, there is the adjoint spinor 

\begin{eqnarray}
\overline{\psi} & = & \left(\begin{array}{cc}
\overline{\psi_{a}} & \overline{\psi}_{b}\end{array}\right)\label{eq:4}\end{eqnarray}
where the spinor field $\psi$, is the described \cite{key-13} as
a quaternion 

\begin{align}
\psi= & \psi_{0}+e_{k}\psi_{k}\,\,\,\,\,(\forall\, k=1,2,3)\label{eq:5}\end{align}
followed by a multiplication rule

\begin{eqnarray}
e_{j}e_{k}=-\delta_{jk}+ & \epsilon_{jkl} & e_{l}\,\,(\forall j,k,l=1,2,3).\label{eq:6}\end{eqnarray}
Here $\delta_{jk}$ and $\epsilon_{jkl}$ are respectively denoted
as Kronecker delta symbol and three index Levi - Civita symbols with
their usual definitions.The quaternion conjugates of quaternion basis
elements as

\begin{eqnarray}
e_{k}^{\dagger} & = & e_{k}.\label{eq:7}\end{eqnarray}
Accordingly the adjoint spinor $\overline{\psi}=$$\psi^{\dagger}\gamma_{0}$
($\psi^{\dagger}$ denotes the Hermitian conjugate spinor) is described
as

\begin{eqnarray}
\overline{\psi}= & \psi_{0}- & e_{k}\psi_{k}\label{eq:8}\end{eqnarray}
whereas a spinor (\ref{eq:3}) is described as a quaternion 

\begin{eqnarray}
\psi=\psi_{0}+e_{1}\psi_{1}+ & e_{2}\psi_{2}+ & e_{3}\psi_{3}\label{eq:9}\end{eqnarray}
which can be decomposed as 

\begin{align}
\psi=( & \psi_{0}+e_{1}\psi_{1})+e_{2}(\psi_{2}-e_{1}\psi_{3})=\psi_{a}+e_{2}\psi_{b}.\label{eq:10}\end{align}
This is the simpletic representation of quaternions in terms of complex
number representations. In equation (\ref{eq:10}), we have written
$\psi_{a}=(\psi_{0}+e_{1}\psi_{1})$ and $\psi_{b}=(\psi_{2}-e_{1}\psi_{3})$
described in terms of the field of real number representations. Accordingly,
we may write 

\begin{equation}
\overline{\psi}=\psi_{0}-e_{1}\psi_{1}-e_{2}\psi_{2}-e_{3}\psi_{3}=(\psi_{0}-e_{1}\psi_{1})-e_{2}\,(\psi_{2}-e_{1}\psi_{3})=\psi_{a}^{\dagger}-e_{2}\psi_{b}^{\dagger}.\label{eq:11}\end{equation}
So, we may write the quaternionic form of the Lagrangian in terms
of $\psi$as 

\begin{eqnarray}
L= & [i\overline{\psi}\gamma^{\mu}\partial_{\mu}\psi & -m\,\overline{\psi}\psi]\label{eq:12}\end{eqnarray}
where $m=\left[\begin{array}{cc}
m_{1} & 0\\
o & m_{2}\end{array}\right]$ is the mass matrix with $m_{1}$ is the mass of the field $\psi_{1}$
whereas $m_{2}$ is that of the field $\psi_{2}$.

\section{Quaternionic Dirac Equation }

Substituting the values of $\psi$ and $\overline{\psi}$ from equations
(\ref{eq:3}) and (\ref{eq:4}) in equation (\ref{eq:12}), we get 

\begin{eqnarray}
L=i\left(\begin{array}{cc}
\overline{\psi}_{a} & \overline{\psi}_{b}\end{array}\right)\gamma^{\mu}\partial_{\mu}\left(\begin{array}{c}
\psi_{a}\\
\psi_{b}\end{array}\right) & -m\left(\begin{array}{cc}
\overline{\psi}_{a} & \overline{\psi}_{b}\end{array}\right) & \left(\begin{array}{c}
\psi_{a}\\
\psi_{b}\end{array}\right)\label{eq:13}\end{eqnarray}
which is reduced to equation (\ref{eq:1}). Defining the Eular Lagrangian
equation as 

\begin{eqnarray}
\partial_{\mu}\left(\frac{\partial L}{\partial(\partial_{\mu}\psi)}\right) & = & \frac{\partial L}{\partial\psi}\label{eq:14}\end{eqnarray}

and taking the variation with respect to $\overline{\psi_{a}}$ and
$\overline{\psi_{b}}$, we get

\begin{eqnarray}
i\gamma^{\mu}(\partial_{\mu}\psi_{a})-m\,\psi_{a} & = & 0\label{eq:15}\end{eqnarray}
and

\begin{eqnarray}
i\gamma^{\mu}(\partial_{\mu}\psi_{b})-m\,\psi_{b} & = & 0.\label{eq:16}\end{eqnarray}
Equations (\ref{eq:15}) and (\ref{eq:16}) are respectively recalled
as the Dirac equations \cite{key-13} for the spinors $\psi_{a}$and
$\psi_{b}$. Similarly if we take the variations with respect to $\psi_{a}$
and $\psi_{b}$ we get 

\begin{eqnarray}
i\,(\partial_{\mu}\overline{\psi}_{a})\gamma^{\mu} & +m\overline{\psi}_{a} & =0\label{eq:17}\end{eqnarray}
and

\begin{eqnarray}
i\,(\partial_{\mu}\overline{\psi}_{b})\gamma^{\mu} & +m\overline{\psi}_{b} & =0\label{eq:18}\end{eqnarray}
which are respectively recalled as the Dirac equations for the adjoint
spinors $\overline{\psi_{a}}$ and $\overline{\psi_{b}}$. In equations
(\ref{eq:15}-\ref{eq:18}) the $\gamma$ matrices are quaternion
valued \cite{key-13} i.e.

\begin{align}
\gamma^{0}=\left[\begin{array}{cc}
1 & 0\\
0 & -1\end{array}\right];\,\,\,\,\, & \gamma^{j}=\left[\begin{array}{cc}
0 & i\, e_{j}\\
-i\, e_{j} & 0\end{array}\right]\,\,(\forall j=1,2,3;\,\,\, i=\sqrt{-1}).\label{eq:19}\end{align}

These $\gamma$ matrices satisfy the following relations

\begin{eqnarray}
\gamma_{\mu}\gamma_{\nu}+\gamma_{\nu}\gamma_{\mu}= & -2 & g_{\mu\nu}\label{eq:20}\end{eqnarray}
where 

\begin{eqnarray}
g_{\mu\nu} & = & (-1,+1,+1,+1)\,\,(\forall\,\,\mu,\,\nu=0,1,2,3).\label{eq:21}\end{eqnarray}
Let us write the Dirac equation in terms of a quaternion valued spinor
$\psi$. Now multiplying equation (\ref{eq:16}) by quaternion basis
element $e_{2}$ , adding the resultant to equation (\ref{eq:15})
and using equation (\ref{eq:10}) , we get the Dirac equation as 

\begin{eqnarray}
i\gamma^{\mu}(\partial_{\mu}\psi)-m\psi & = & 0.\label{eq:22}\end{eqnarray}
 Similarly, we may write the quaternion conjugate Dirac equation as 

\begin{eqnarray}
i(\partial_{\mu}\overline{\psi})\gamma^{\mu}+m\overline{\psi} & = & 0.\label{eq:23}\end{eqnarray}
Dirac equations (\ref{eq:22}-\ref{eq:23}) provide the four current
as 

\begin{eqnarray}
j^{\mu} & = & \overline{\psi}\,\gamma^{\mu}\psi\label{eq:24}\end{eqnarray}
which satisfies the continuity equation $\partial_{\mu}j^{\mu}=0$.

\section{Quaternionic $SU(2)$ Global gauge symmetry}

In global gauge symmetry , the unitary transformations are independent
of space and time. Accordingly, under $SU(2)$ global gauge symmetry,
the quaternion spinor $\psi$ transforms as 

\begin{eqnarray}
\psi & \longmapsto\psi^{\shortmid}= & U\,\psi\label{eq:25}\end{eqnarray}
where $U$ is $2\times2$ unitary matrix and satisfies 

\begin{eqnarray}
U^{\dagger}U & =UU^{\dagger}=UU^{-1}=U^{-1}U= & 1.\label{eq:26}\end{eqnarray}
On the other hand, the quaternion conjugate spinor transforms as \begin{eqnarray}
\overline{\psi}\longmapsto\overline{\psi^{\shortmid}} & = & \overline{\psi}U^{-1}\label{eq:27}\end{eqnarray}
and hence the combination $\psi\overline{\psi}=\overline{\psi}\psi=\psi\overline{\psi^{\shortmid}}=\overline{\psi^{\shortmid}}\psi$
is an invariant quantity. We may thus write any unitary matrix as 

\begin{eqnarray}
U & = & \exp\left(i\,\hat{H}\right)\label{eq:28}\end{eqnarray}
where $H$ is Hermitian $H^{\dagger}=H$. Thus, we may express the
Hermitian $2\times2$ matrix in terms of four real numbers, $a_{1,}a_{2},\, a_{3},$
and $\theta$ as 

\begin{eqnarray}
\hat{H} & = & \theta1+\sigma_{j}a_{j}=\theta1+ie_{j}a_{j}\label{eq:29}\end{eqnarray}
where $1$ is the $2\times2$ unit matrix, $\sigma_{j}$ are well
known $2\times2$ Pauli-spin matrices and $e_{1},\, e_{2},\, e_{3}$
are the quaternion units which are connected with Pauli-spin matrices
as \begin{eqnarray}
e_{0}= & 1;\,\,\,\,\, & e_{j}=-i\sigma_{j}.\label{eq:30}\end{eqnarray}
 Hence, we write the Hermitian matrix $H$ as 

\begin{eqnarray}
H & = & \left(\begin{array}{cc}
\theta+a_{3} & a_{1}-ia_{2}\\
a_{1}+ia_{2} & \theta-a_{3}\end{array}\right).\label{eq:31}\end{eqnarray}

Equation (\ref{eq:28}) may now be reduced as 

\begin{eqnarray}
U & = & \exp\left(i\,\theta\right).\exp\left(-e_{j}a_{j}\right).\label{eq:32}\end{eqnarray}
For $SU(2)$ global gauge transformations both $\theta$ and $\overrightarrow{a}$
are independent of space time. Here $\exp\left(i\,\theta\right)$
describes the $U(1)$ gauge transformation while the term $\exp\left(-e_{j}a_{j}\right)$
represents the non-Abelian $SU(2)$ gauge transformations. Thus under
global $SU(2)$ gauge transformations, the Dirac spinor $\psi$ transforms
as 

\begin{eqnarray}
\psi & \longmapsto\psi^{\shortmid}= & U\,\psi=\exp\left(-e_{j}a_{j}\right)\psi.\label{eq:33}\end{eqnarray}
The generators of this group $e_{i}$ obey the commutation relation;

\begin{eqnarray}
\left[e_{j},e_{k}\right] & = & 2f_{jkl}e_{l}\label{eq:34}\end{eqnarray}
which implies $e_{i}e_{j}\neq e_{j}e_{i}$ showing that the elements
of the group are not commutating giving rise to the non abelian gauge
structure. So, the partial derivative of spinor $\psi$accordingly
transforms as 

\begin{eqnarray}
\partial_{\mu}\psi(x) & \longmapsto & \partial_{\mu}\psi^{'}(x)=\exp\left(-e_{j}a_{j}\right)(\partial_{\mu}\psi).\label{eq:35}\end{eqnarray}
As such the Lagrangian density is invariant under $SU(2$) global
gauge transformations i.e. $\delta L=0.$ The Lagrangian density thus
yields the continuity equation after taking the variations and the
definitions of Euler Lagrange equations as 

\begin{eqnarray}
\partial_{\mu}\left\{ \frac{\partial L}{\partial(\partial_{\mu}\psi)}e_{k}\psi\right\}  & = & \partial_{\mu}\left\{ i\overline{\psi}\gamma^{\mu}e_{k}\psi\right\} =\partial_{\mu}(j^{\mu})^{k}=0\,\,(\forall\, k=1,2,3)\label{eq:36}\end{eqnarray}
where the $SU(2)$ gauge current is defined as 

\begin{eqnarray}
(j^{\mu})^{k} & = & \left\{ i\overline{\psi}\gamma^{\mu}e_{k}\psi\right\} .\label{eq:37}\end{eqnarray}

which is the global current of the fermion field.

\section{Quaternionic $SU(2)$ Local Gauge Symmetry }

For $SU(2)$ local gauge transformation we may replace the unitary
gauge transformation as space- time depaendent. So replacing $U$
by $S$ in equation (\ref{eq:25}), we get

\begin{eqnarray}
\psi & \longmapsto\psi^{\shortmid}= & S\,\psi\label{eq:38}\end{eqnarray}
in which \begin{eqnarray}
S & = & \exp[-\sum_{j}\, q\, e_{j}\zeta_{j}(x)]\label{eq:39}\end{eqnarray}
where parameter $\overrightarrow{\zeta}=-\frac{\overrightarrow{a}(x)}{q}$
with $\overrightarrow{a}(x)$ is infinitesimal quantity depending
on space and time and $q$ is described as the coupling constant.
Consequently, the Lagrangian density (\ref{eq:13}) is no more invariant
under $SU(2)$ local gauge symmetry as the partial derivative picks
an extra term i.e. 

\begin{eqnarray}
\partial_{\mu}\psi\mapsto S\partial_{\mu}\psi & +\left(\partial_{\mu}S\right)\psi & =D_{\mu}\psi\label{eq:40}\end{eqnarray}
where the covariant derivative $D_{\mu}$ has been defined in terms
of two $Q-$ gauge fields i.e 

\begin{eqnarray}
D_{\mu}\psi & =\partial_{\mu}\psi & +A_{µ}\psi+B_{µ}\psi\label{eq:41}\end{eqnarray}
where $A_{\mu}=-iA_{\mu}^{j}\sigma_{j}=A_{\mu}^{j}e_{j}=\overrightarrow{A_{\mu}}\cdot\overrightarrow{e}$
and $B_{\mu}=-iB_{\mu}^{j}\sigma_{j}=B_{\mu}^{j}e_{j}=\overrightarrow{B_{\mu}}\cdot\overrightarrow{e}$.
Two gauge fields $A_{\mu}$ and $B_{\mu}$are respectively associated
with electric and magnetic charges of dyons (i.e particles carrying
the simultaneous existence of electric and magnetic charges). Thus
the gauge field $\left\{ A_{\mu}\right\} $ is coupled with the electric
charge while the gauge field $\left\{ B_{\mu}\right\} $ is coupled
with the magnetic charge (i.e. magentic monopole). These two gauge
fields are subjected by the following gauge transformations

\begin{equation}
A_{µ}'\longmapsto S\, A_{µ}\, S^{-1}+(\partial_{\mu}S)\, S^{-1};\,\,\,\,\: B_{µ}'\longmapsto S\, B_{µ}\, S^{-1}+(\partial_{\mu}S)\, S^{-1}.\label{eq:42}\end{equation}
For the limiting case of infinitesimal transformations of $\zeta$
, we may expand $S$ by keeping only first order terms as

\begin{equation}
S\cong1+\overrightarrow{e}.\overrightarrow{a}(x);\,\,\,\,\,\,\, S^{-1}\cong1-\overrightarrow{e}.\overrightarrow{a}(x);\,\,\,\:\partial_{\mu}\left(S\right)\cong\overrightarrow{e}.\partial_{\mu}\left\{ \overrightarrow{a}\left(x\right)\right\} .\label{eq:43}\end{equation}
So, on replacing partial derivative of global gauge symmetry to covariant
derivative of local gauge symmetry, we may write the invariant Lagrangian
density for the quaternion $SU(2)$ gauge fields in the following
form\begin{eqnarray}
L & = & i\overline{\psi}\gamma_{\mu}(D_{\mu}\psi)-m\overline{\psi}\psi,\label{eq:44}\end{eqnarray}
which yields the following current densities of electric and magnetic
charges of dyons i.e

\begin{equation}
J_{\mu}=\left(j_{\mu}\right)_{electric}+\left(j_{\mu}\right)_{magnetic}=i\mathtt{e}\overline{\psi}\,\gamma_{\mu}\psi+i\mathtt{g}\overline{\psi}\,\gamma_{\mu}\psi.\label{eq:45}\end{equation}
where $\mathtt{e}$ is the electric charge and $\mathtt{g}$ is the
magnetic charge. Equation (\ref{eq:45}) does not satisfy the usual
continuity equation i.e. $\partial^{\mu}J_{\mu}\neq0$ but satisfies
the Noetherian form of continuity equation with covariant derivative
as

\begin{equation}
D_{\mu}J^{\mu}=0.\label{eq:46}\end{equation}

\section{Definition of Octonions }

An octonion $x$ is expressed \cite{key-15} as a set of eight real
numbers

\begin{equation}
x=e_{0}x_{0}+e_{1}x_{1}+e_{2}x_{2}+e_{3}x_{3}+e_{4}x_{4}+e_{5}x_{5}+e_{6}x_{6}+e_{7}x_{7}=e_{0}x_{0}+\sum_{A=1}^{7}e_{A}x_{A}\label{eq:47}\end{equation}
where $e_{A}(A=1,2,...,7)$ are imaginary octonion units and $e_{0}$
is the multiplicative unit element. Set of octets $(e_{0},\, e_{1},\, e_{2},e_{3},e_{4},e_{5},e_{6},e_{7})$
are known as the octonion basis elements and satisfy the following
multiplication rules

\begin{equation}
e_{0}=1;\,\, e_{0}e_{A}=e_{A}e_{0}=e_{A};\,\,\,\, e_{A}e_{B}=-\delta_{AB}e_{0}+f_{ABC}e_{C}.\,\,(A,B,C=1,2,.....,7)\label{eq:48}\end{equation}
The structure constants $f_{ABC}$ is completely antisymmetric and
takes the value $1$ for following combinations,

\begin{eqnarray}
f_{ABC}= & +1; & \forall(ABC)=(123),\,(471),\,(257),\,(165),\,(624),\,(543),\,(736).\label{eq:49}\end{eqnarray}
It is to be noted that the summation convention is used for repeated
indices. Here the octonion algebra $\mathcal{O}$ is described over
the algebra of real numbers having the vector space of dimension $8$.
As such we may write the following relations among octonion basis
elements 

\begin{eqnarray}
\left[e_{A},\,\, e_{B}\right] & = & 2\, f_{ABC}\, e_{C};\nonumber \\
\left\{ e_{A},\,\, e_{B}\right\}  & = & -2\,\delta_{AB}e_{0};\nonumber \\
e_{A}(\, e_{B}\, e_{C}) & \neq & (e_{A}\, e_{B}\,)\, e_{C}\label{eq:50}\end{eqnarray}
where brackets $[\,\,,\,\,]$ and $\{\,\,,\,\,\}$ are used respectively
for commutation and the anti commutation relations while $\delta_{AB}$
is the usual Kronecker delta-Dirac symbol. Octonion conjugate is defined
as

\begin{equation}
\overline{x}=e_{0}x_{0}-e_{1}x_{1}-e_{2}x_{2}-e_{3}x_{3}-e_{4}x_{4}-e_{5}x_{5}-e_{6}x_{6}-e_{7}x_{7}=e_{0}x_{0}-\sum_{A=1}^{7}e_{A}x_{A}\label{eq:51}\end{equation}
where we have used the conjugates of basis elements as $\overline{e_{0}}=e_{0}$
and $\overline{e_{A}}=-e_{A}$. Hence an octonion can be decomposed
in terms of its scalar $(Sc(x))$ and vector $(Vec(x))$ parts as 

\begin{eqnarray}
Sc(x) & = & \frac{1}{2}(\, x\,+\,\overline{x}\,);\,\,\,\,\,\, Vec(x)=\frac{1}{2}(\, x\,-\,\overline{x}\,)=\sum_{A=1}^{7}\, e_{A}x_{A}.\label{eq:52}\end{eqnarray}
Conjugates of product of two octonions and its own are described as

\begin{eqnarray}
\overline{(x\, y)}= & \overline{y}\,\,\overline{x};\,\,\,\,\,\, & \overline{(\overline{x})}=x.\label{eq:53}\end{eqnarray}
while the scalar product of two octonions is defined as 

\begin{eqnarray}
\left\langle x\,,\, y\right\rangle  & =\frac{1}{2}(x\,\overline{y}+y\,\overline{x})=\frac{1}{2}(\overline{x}\, y+\overline{y}\, x)= & \sum_{\alpha=0}^{7}\, x_{\alpha}\, y_{\alpha}.\label{eq:54}\end{eqnarray}
The norm $N(x)$ and inverse $x^{-1}$(for a nonzero $x$) of an octonion
are respectively defined as

\begin{eqnarray}
N(x)=x\,\overline{x}=\overline{x}\, x & = & \sum_{\alpha=0}^{7}\, x_{\alpha}^{2}.e_{0};\nonumber \\
x^{-1} & = & \frac{\overline{x}}{N(x)}\,\Longrightarrow x\, x^{-1}=x^{-1}\, x=1.\label{eq:55}\end{eqnarray}
The norm $N(x)$ of an octonion $x$ is zero if $x=0$, and is always
positive otherwise. It also satisfies the following property of normed
algebra

\begin{eqnarray}
N(x\, y)= & N(x)\, N(y)= & N(y)\, N(x).\label{eq:56}\end{eqnarray}
Equation (\ref{eq:50}) shows that octonions are not associative in
nature and thus do not form the group in their usual form. Non - associativity
of octonion algebra $\mathcal{O}$ is provided by the associator $(x,y,z)=(xy)z-x(yz)\,\,\forall x,y,z\in\mathcal{O}$
defined for any three octonions. If the associator is totally antisymmetric
for exchanges of any three variables, i.e. $(x,y,z)=-(z,y,x)=-(y,x,z)=-(x,z,y)$,
then the algebra is called alternative. Hence, the octonion algebra
is neither commutative nor associative but, is alternative.

\section{Gellmann $\lambda$ matrices }

In order to extend the symmetry from $SU(2)$ to $SU(3)$ we replace
three Pauli spin matrices by eight Gellmann $\lambda$ matrices. $\lambda_{j}\,\:\left(j=1,2,......8\right)$
be the $3\times3$ traceless Hermitian matrices introduced by Gell-Mann.Their
explicit forms are;

\begin{eqnarray*}
\lambda_{1}=\left[\begin{array}{ccc}
0 & 1 & 0\\
1 & 0 & 0\\
0 & 0 & 0\end{array}\right] & , & \lambda_{2}=\left[\begin{array}{ccc}
0 & -i & 0\\
i & 0 & 0\\
0 & 0 & 0\end{array}\right],\end{eqnarray*}

\begin{eqnarray*}
\lambda_{3}=\left[\begin{array}{ccc}
1 & 0 & 0\\
0 & -1 & 0\\
0 & 0 & 0\end{array}\right] & , & \lambda_{4}=\left[\begin{array}{ccc}
0 & 0 & 1\\
0 & 0 & 0\\
1 & 0 & 0\end{array}\right],\end{eqnarray*}

\begin{eqnarray*}
\lambda_{5}=\left[\begin{array}{ccc}
0 & 0 & -i\\
0 & 0 & 0\\
i & 0 & 0\end{array}\right] & , & \lambda_{6}=\left[\begin{array}{ccc}
0 & 0 & 0\\
0 & 0 & 1\\
0 & 1 & 0\end{array}\right],\end{eqnarray*}

\begin{eqnarray}
\lambda_{7}=\left[\begin{array}{ccc}
0 & 0 & 0\\
0 & 0 & -i\\
0 & i & 0\end{array}\right] & , & \lambda_{8}=\frac{1}{\sqrt{3}}\left[\begin{array}{ccc}
1 & 0 & 0\\
0 & 1 & 0\\
0 & 0 & -2\end{array}\right];\label{eq:57}\end{eqnarray}
which satisfy the following properties as

\begin{align}
\left(\lambda_{j}\right)^{\dagger}= & \lambda_{j};\nonumber \\
Tr\lambda_{j}= & 0;\nonumber \\
Tr\left(\lambda_{j}\lambda_{k}\right)= & 2\delta_{jk};\nonumber \\
\left[\lambda_{j},\lambda_{k}\right] & =2F_{jkl}\lambda_{l}\,\,(\forall\,\, j,k,l=1,2,3,4,5,6,7,8);\label{eq:58}\end{align}
where $F_{jkl}$ are the structure constants of $SU(3)$ group defined
as 

\begin{equation}
F_{123}=1;\,\, F_{147}=F_{257}=F_{435}=F_{651}=F_{637}=\frac{1}{2};\,\, F_{458}=F_{678}=\sqrt{\frac{3}{2}}.\label{eq:59}\end{equation}

\section{Relation between Octonion and Gellmann Matrices}

Let us establish the relationship between octonion basis elements
$e_{A}$ and Gellmann $\lambda$ matrices. Comparing equations (\ref{eq:50})
and (\ref{eq:58}), we get

\begin{eqnarray}
F_{ABC} & = & f_{ABC}\,\,(\forall\, ABC=123)\label{eq:60}\end{eqnarray}
and 

\begin{equation}
F_{ABC}=\frac{1}{2}\, f_{ABC}\,\,(\forall\, ABC=147,246,257,435,516,637).\label{eq:61}\end{equation}
Equation (\ref{eq:60}) leads to 

\begin{equation}
\frac{\left[e_{A},e_{B}\right]}{\left[\lambda_{A},\lambda_{B}\right]}=\frac{e_{C}}{i\lambda_{C}}\,\,(\forall\, A,B,C=1,2,3)\Rightarrow\left[e_{A},e_{B}\right]=\left[\lambda_{A},\lambda_{B}\right]\,\,(\forall\, e_{C}=i\lambda_{C}).\label{eq:62}\end{equation}
On the other hand equation (\ref{eq:61}) gives rise to 

\begin{align}
\frac{\left[e_{A},e_{B}\right]}{\left[\lambda_{A},\lambda_{B}\right]}= & \frac{e_{C}}{2i\lambda_{C}}\,(\forall A,B,C=516,624,471,435,673,572)\nonumber \\
\Rightarrow\left[e_{A},e_{B}\right]= & \left[\lambda_{A},\lambda_{B}\right]\,\,\,(\forall\, e_{C}=i\frac{\lambda_{C}}{2}).\label{eq:63}\end{align}
We may now describe the correspondence between the matrix $\lambda_{8}$
and octonion units in the following manner i.e. 

\begin{equation}
\lambda_{8}\Longrightarrow-\frac{2}{i\sqrt{3}}\left\{ \left[e_{4},e_{5}\right]+\left[e_{6},e_{7}\right]\right\} \Longrightarrow-\frac{2}{i\sqrt{3}}\left(e_{4}e_{5}-e_{5}e_{4}+e_{6}e_{7}-e_{7}e_{6}\right)\Longrightarrow\frac{8e_{3}}{i\sqrt{3}}.\label{eq:64}\end{equation}
Hence we may describe one to mapping (interrelationship) between octonion
basis elements and Gellmann $\lambda$ matrices by using equations
(\ref{eq:62}-\ref{eq:63}) as,

\begin{equation}
e_{A}\propto\lambda_{A}\Rightarrow e_{A}=k\lambda_{A}\label{eq:65}\end{equation}
where $k$ is proportionality constant depending on the different
values of $A$ i.e.$k=i\,(\forall\, A=1,2,3)$ and $k=\frac{i}{2}\,(\forall\, ABC=516,624,471,435,673,572).$
From equation (\ref{eq:64}), we also get

\begin{equation}
e_{3}\propto\lambda_{8}\Rightarrow e_{3}=\mathbb{\mathtt{k}}\lambda_{8}\label{eq:66}\end{equation}
where $\mathtt{k}=\frac{i\sqrt{3}}{8}$.With these relations between
the octonion units and Gellmann $\lambda$ matrices, we may develop
the octonion quantum chromodynamics in consistent way. To do this,
let us establish the following commutation relations for octonion
basis elements and Gellmann $\lambda$ matrices i.e. 

\begin{align}
\frac{\left[e_{6},e_{5}\right]}{\left[\lambda_{6},\lambda_{5}\right]}= & \frac{\left[e_{4},e_{7}\right]}{\left[\lambda_{4},\lambda_{7}\right]}=\frac{e_{1}}{2i\lambda_{1}}=\frac{1}{2k_{1}}\Longrightarrow\lambda_{1}=-ie_{1}k_{1};\nonumber \\
\frac{\left[e_{4},e_{6}\right]}{\left[\lambda_{4},\lambda_{6}\right]} & =\frac{\left[e_{5},e_{7}\right]}{\left[\lambda_{5},\lambda_{7}\right]}=\frac{e_{2}}{2i\lambda_{2}}=\frac{1}{2k_{2}}\Longrightarrow\lambda_{2}=-ie_{2}k_{2};\nonumber \\
\frac{\left[e_{5},e_{4}\right]}{\left[\lambda_{5},\lambda_{4}\right]} & =\frac{\left[e_{6},e_{7}\right]}{\left[\lambda_{6},\lambda_{7}\right]}=\frac{e_{3}}{2i\lambda_{3}}=\frac{1}{2k_{3}}\Longrightarrow\lambda_{3}=-ie_{3}k_{3};\nonumber \\
\frac{\left[e_{7},e_{1}\right]}{\left[\lambda_{7},\lambda_{1}\right]} & =\frac{\left[e_{6},e_{2}\right]}{\left[\lambda_{6},\lambda_{2}\right]}=\frac{\left[e_{3},e_{5}\right]}{\left[\lambda_{3},\lambda_{5}\right]}=\frac{e_{4}}{2i\lambda_{4}}=\frac{1}{2k_{4}}\Longrightarrow\lambda_{4}=-ie_{4}k_{4};\nonumber \\
\frac{\left[e_{4},e_{3}\right]}{\left[\lambda_{4},\lambda_{3}\right]} & =\frac{\left[e_{7},e_{2}\right]}{\left[\lambda_{7},\lambda_{2}\right]}=\frac{\left[e_{1},e_{6}\right]}{\left[\lambda_{1},\lambda_{6}\right]}=\frac{e_{5}}{2i\lambda_{5}}=\frac{1}{2k_{5}}\Longrightarrow\lambda_{5}=-ie_{5}k_{5};\nonumber \\
\frac{\left[e_{5},e_{1}\right]}{\left[\lambda_{5},\lambda_{1}\right]} & =\frac{\left[e_{7},e_{3}\right]}{\left[\lambda_{7},\lambda_{3}\right]}=\frac{\left[e_{2},e_{4}\right]}{\left[\lambda_{2},\lambda_{4}\right]}=\frac{e_{6}}{2i\lambda_{6}}=\frac{1}{2k_{6}}\Longrightarrow\lambda_{6}=-ie_{6}k_{6};\nonumber \\
\frac{\left[e_{1},e_{4}\right]}{\left[\lambda_{1},\lambda_{4}\right]} & =\frac{\left[e_{2},e_{5}\right]}{\left[\lambda_{2},\lambda_{5}\right]}=\frac{\left[e_{3},e_{6}\right]}{\left[\lambda_{3},\lambda_{6}\right]}=\frac{e_{7}}{2i\lambda_{7}}=\frac{1}{2k_{7}}\Longrightarrow\lambda_{7}=-ie_{7}k_{7}.\label{eq:67}\end{align}
As such, we may get the following relationship between Gell Mann $\lambda$matrices
and octonion units:

\begin{align}
\lambda_{A}= & -i\, e_{A}k_{A}\,\,(\forall A=1,2,3,4,5,6,7);\nonumber \\
\lambda_{8}= & -i\,\frac{8}{\sqrt{3}}\, e_{3}.\label{eq:68}\end{align}

\section{Octonionic Reformulation of QCD}

The local gauge theory of color $SU(3)$group gives the theory of
QCD. The QCD (quantum chromodynamics) is very close to Yang-Mills
(non Abelian) gauge theory. The above mentioned $SU(2)$ gauge symmetry
describes the symmetry of the weak interactions. On the other hand,
the theory of strong interactions,quantum chromodynamics (QCD), is
based on colour $SU(3)$ ( namely $SU(3)_{c}$) group. This is a group
which acts on the colour indices of quark favours described in the
form of a basic triplet i.e.

\begin{eqnarray}
\psi & = & \left(\begin{array}{c}
\psi_{1}\\
\psi_{2}\\
\psi_{3}\end{array}\right)\rightarrow\left(\begin{array}{c}
R\\
B\\
G\end{array}\right)\label{eq:69}\end{eqnarray}
where indices $R$, $B$, and $G$ are the three colour of quark flavours.
Under $SU(3)_{c}$symmetry, the spinor $\psi$ transforms as 

\begin{align}
\psi\longmapsto & \psi^{\shortmid}=U\psi=\exp\left\{ i\lambda_{a}\alpha^{a}(x)\right\} \psi\label{eq:70}\end{align}
where $\lambda$ are Gellmann matrices , $a=1,2,......8$ and the
parameter $\alpha$ is space time dependent. We may develop accordingly
the octonionic reformulation of quantum chromodynamics (QCD) on replacing
the Gellmann $\lambda$ matrices by octonion basis elements $e_{A}$
given by equations (\ref{eq:65}) and (\ref{eq:66}). Now calculating
the value of $\lambda_{a}\alpha^{a}\left(x\right)=\sum_{a=1}^{8}\lambda_{a}\alpha^{a}\left(x\right)$
and using the relations between GellMann $\lambda$ matrices and octonion
units given by equations\textbf{ $\left(\ref{eq:68}\right)$}, we
find

\begin{align}
\sum_{a=1}^{8}\lambda_{a}\alpha^{a}\left(x\right) & =-ie_{1}k_{1}\alpha^{1}\left(x\right)-ie_{2}k_{2}\alpha^{2}\left(x\right)-ie_{3}\left(k_{3}\alpha^{3}\left(x\right)+k_{8}\alpha^{8}\left(x\right)\right)\nonumber \\
- & ie_{4}k_{4}\alpha^{4}\left(x\right)-ie_{5}k_{5}\alpha^{5}\left(x\right)-ie_{6}k_{6}\alpha^{6}\left(x\right)-ie_{7}k_{7}\alpha^{7}\left(x\right).\label{eq:71}\end{align}
Now taking following transformations

\begin{align}
k_{1}\alpha_{1}\longmapsto & \beta^{1}\,;\, k_{2}\alpha_{2}\longmapsto\beta^{2}\,;\,\left(k_{3}\alpha_{3}+k_{8}\alpha_{8}\right)\longmapsto\beta^{3};\nonumber \\
k_{4}\alpha_{4} & \longmapsto\beta^{4}\,;\, k_{5}\alpha_{5}\longmapsto\beta^{5}\,;\, k_{6}\alpha_{6}\longmapsto\beta^{6}\,;\, k_{7}\alpha_{7}\longmapsto\beta^{7};\label{eq:72}\end{align}
we get 

\begin{equation}
\sum_{a=1}^{8}\lambda_{a}\alpha^{a}\left(x\right)=-ie_{1}\beta^{1}-ie_{2}\beta^{2}-ie_{3}\beta^{3}-ie_{4}\beta^{4}-ie_{5}\beta^{5}-ie_{6}\beta^{6}-ie_{7}\beta^{7}.\label{eq:73}\end{equation}
It may also be written in the following generalized compact form i.e. 

\begin{equation}
\sum_{a=1}^{8}\lambda_{a}\alpha^{a}\left(x\right)=-i\sum_{q=1}^{7}e_{q}\beta^{q}\left(x\right);\label{eq:74}\end{equation}
which may be written in terms of the following traceless Hermitian
matrix form as

\begin{equation}
-i\sum_{q=1}^{7}e_{q}\beta^{q}\left(x\right)=\left[\begin{array}{ccc}
\alpha_{3}+\frac{\alpha_{8}}{\sqrt{3}} & \alpha_{1}-i\alpha_{2} & \alpha_{4}-i\alpha_{5}\\
\alpha_{1}+i\alpha_{2} & -\alpha_{3}+\frac{\alpha_{8}}{\sqrt{3}} & \alpha_{6}-i\alpha_{7}\\
\alpha_{4}+i\alpha_{5} & \alpha_{6}+i\alpha_{7} & -\frac{2\alpha_{8}}{\sqrt{3}}\end{array}\right].\label{eq:75}\end{equation}
Now $\left(\ref{eq:69}\right)$ becomes

\begin{equation}
\psi\longmapsto\psi^{\shortmid}=U\psi=\exp\left\{ e_{q}\beta^{q}(x)\right\} \label{eq:76}\end{equation}
So we may write the locally gauge invariant $SU(3)_{c}$ ,Lagrangian
density in the following form;\begin{eqnarray}
L_{local}= & \left(i\overline{\psi}\gamma_{\mu}D_{\mu}\psi-m\overline{\psi}\psi\right) & -\frac{1}{4}G_{\mu\nu}^{a}G_{a}^{\mu\nu}\label{eq:77}\end{eqnarray}
where 

\begin{eqnarray}
D_{\mu}\psi & = & \partial_{\mu}\psi+\mathsf{\boldsymbol{e}\,}e_{a}A_{\mu}^{a}\psi+\mathsf{g\,}e_{a}B_{\mu}^{a}\psi\label{eq:78}\end{eqnarray}
and 

\begin{align}
G_{\mu\nu}^{a}= & \partial_{\mu}V_{\nu}^{a}-\partial_{\nu}V_{\mu}^{a}-\mathbf{q\,}f_{abc}V_{\mu}^{b}V_{\nu}^{c}\nonumber \\
= & \left(\partial_{\mu}A_{\nu}^{a}-\partial_{\nu}A_{\mu}^{a}-\mathbf{\mathbf{e}\,}f_{abc}A_{\mu}^{b}A_{\nu}^{c}\right)+\left(\partial_{\mu}B_{\nu}^{a}-\partial_{\nu}B_{\mu}^{a}-\mathbf{\mathbf{g}\,}f_{abc}B_{\mu}^{b}B_{\nu}^{c}\right).\label{eq:79}\end{align}
Here in equations (\ref{eq:78}) and (\ref{eq:79}), the $\mathbf{e}$
and $\mathbf{g}$ are the coupling constants due to the occurence
of respectively the electric and magnetic charges on dyons. On the
similar ground the two gauge fields $\left\{ A_{\mu}\right\} $and
$\left\{ B_{\mu}\right\} $ are present in the theory due to the occurence
of respectively the electric and magnetic charges on dyons. As such,
in the present theory we have two kinds of color gauge groups respectively
associated with the two gauge fields of electric and magnetic charges
on dyons. Hence the locally gauge covariant Lagrangian density is
written as 

\begin{align}
L_{local}= & \left(i\overline{\psi}\gamma_{\mu}\partial_{\mu}\psi-m\overline{\psi}\psi\right)-\mathbf{e\,}\left(\overline{\psi}\gamma^{\mu}\psi\right)e_{a}A_{\mu}^{a}-\mathbf{g\,}\left(\overline{\psi}\gamma^{\mu}\psi\right)e_{a}B_{\mu}^{a}-\frac{1}{4}G_{\mu\nu}^{a}G_{a}^{\mu\nu}\label{eq:80}\end{align}
which leads to the folowing expression for the gauge covariant current
density of coloured dyons

\begin{equation}
J_{\mu}^{a}=\mathbf{e\,}\left(\overline{\psi}\gamma^{\mu}\psi\right)e_{a}+\mathbf{g\,}\left(\overline{\psi}\gamma^{\mu}\psi\right)e_{a}.\label{eq:81}\end{equation}
which leads to the conservation of Noetherian current in octonion
formulation of $SU(3)_{c}$ gauge theory of quantum chromodynamics
(QCD) i.e.

\begin{align}
D_{\mu}J^{\mu} & =0\label{eq:82}\end{align}
where $J^{\mu}=J^{\mu a}\lambda_{a}.$

\end{document}